\documentclass[journal]{IEEEtran2012}

\IEEEoverridecommandlockouts  %for \thanks{}Use \thanks{} after title one for
\usepackage{cite,amssymb}
\usepackage[capitalise]{cleveref}  %For a great convenience in \ref{} usage.
\usepackage[cmex10]{amsmath}
\usepackage[boxed,linesnumbered,vlined,shortend]{algorithm2e}
\usepackage{graphicx}
\usepackage[caption=false,font=footnotesize]{subfig}
\usepackage{url}
\usepackage{placeins} %for \FloatBarrier
\usepackage{enumitem} \setlist{nolistsep}
\usepackage{xcolor}
\usepackage{soul} %soul: for \hl{..} and \sethlcolor{..} commands used 
\definecolor{lightblue}{rgb}{0.85,0.85,1}
\sethlcolor{lightblue} %Frm "soul" package.'yellow' is default & 
\newcommand{\subparagraph}{}  \usepackage[compact]{titlesec} %for section
\usepackage{etoolbox} %for undef used for algorithm2e's caption 
\DeclareMathOperator{\AND}{and}

\DeclareMathOperator{\NaN}{NaN}

\titlespacing*{\section}{0pt}{2pt}{2pt}
\titlespacing*{\subsection}{0pt}{2pt}{1pt}
\renewcommand{\b}[1]{\mathbf{#1}}

\makeatletter
\renewcommand{\algocf@caption@boxruled}{%
  \hrule
  \hbox to \hsize{%
    \vrule\hskip-0.4pt
    \vbox{
       \vskip\interspacetitleboxruled%
       \unhbox\algocf@capbox\hfill
       \vskip\interspacetitleboxruled
       }%
     \hskip-0.4pt\vrule%
   }\nointerlineskip%
}%
\undef{\@algocf@capt@boxed}
\def\@algocf@capt@boxed{above}%%
\makeatother
\usepackage{tikz}
\usetikzlibrary{arrows,calc}

 \renewcommand{\O}[1]{\mathcal{O}(#1)}

 \renewcommand{\b}[1]{\mathbf{#1}}
\begin{document}

% \title{A Comparative Study of Polar Code Constructions}
% \title{Achieving the Best Performance with \\Any Polar Code Construction in AWGN 
% Channel}
% \title{On Polar Code Constructions for the AWGN Channel}
\title{A Comparative Study of Polar Code Constructions for the AWGN Channel}

\author{Harish~Vangala,~Emanuele~Viterbo~and~Yi~Hong \thanks{Authors 
are with the Department of Electrical and Computer Systems Engineering, Monash 
University, Melbourne, VIC 3800, Australia. Email:~\{harish.vangala, 
emanuele.viterbo, yi.hong\}@monash.edu.\vspace{1mm}}
\thanks{This work is supported by NPRP grant \#NPRP5-597-2-241  from the Qatar 
National Research Fund (a member of Qatar Foundation).} \vspace{-1cm}}

\markboth{DRAFT SUBMISSION}{}
% {Journal of Communications and Networks, Vol. XX, No. XX, MMMM, 20XX}{}

\maketitle

\begin{abstract}
\nocite{ARI09tit,ARI08clett,TAL13,TRI12,MOR09clett,MOR09isit,ZHA14clett,LI13,WU14, 
KER14,BON12arxiv2,ZHA11}
We present a comparative study of the performance of various polar code constructions 
in an additive white Gaussian noise (AWGN) channel. A polar code construction is any 
algorithm that selects $K$ best among $N$ possible polar bit-channels at the 
\emph{design signal-to-noise-ratio} (design-SNR) in terms of bit error rate (BER). 
Optimal polar code construction is hard and therefore many suboptimal polar code 
constructions have been proposed at different computational complexities. Polar codes 
are also \emph{non-universal} meaning the code changes significantly with the 
design-SNR. However, it is not known which construction algorithm at what design-SNR 
constructs the best polar codes. We first present a comprehensive survey of all the 
well-known polar code constructions along with their full implementations. We then 
propose a heuristic algorithm to find the best design-SNR for constructing best 
possible polar codes from a given construction algorithm. The proposed algorithm 
involves a search among several possible design-SNRs. We finally use our algorithm 
to perform a comparison of different construction algorithms using extensive 
simulations. We find that all polar code construction algorithms generate equally 
good polar codes in an AWGN channel, if the design-SNR is optimized.
\end{abstract}


\section{Introduction}
Polar codes have been the subject of active research in recent times, mainly due to 
the fact that they are the first ever \emph{provably} capacity achieving codes, with 
explicit construction and very low complexity of encoding and decoding. The polar 
codes were invented by Erdal Arikan \cite{ARI09tit}, using a novel concept called 
\emph{channel polarization}. Soon after, both the concept of channel polarization as 
well as polar codes have been extended to a number of applications and 
generalizations \cite{SAS11book, ABB11ita, CRO10, ARI10isit, MOR14, SAH11allerton, 
SAH13, PAR13tit, BRA13, YAN12, YAN13, SI13, AKU14, SAT10tit, SAS14}.

Let us consider a binary input discrete memoryless symmetric (BI-DMS) channel.
Channel polarization is a technique by which one manufactures $N$ polarized channels 
(called \emph{bit-channels}) out of $N$ identical independent copies of BI-DMS 
channels. The channels are \emph{polarized} without any loss of capacity, in the 
sense that they are either extremely noisy or noiseless as $N\to \infty$.  Then one 
can easily achieve a rate of transmission close to capacity, simply by choosing to 
transmit over only the good bit-channels. However, at any finite blocklength $N$ and 
rate $R\triangleq K/N$, a ranking algorithm for the bit-channels according to 
their bit error rate (BER) becomes necessary to select $K$ good channels out of $N$. 
Here, $K$ is the number of information bits in each code word of length 
$N$. This selection of bit-channels completely defines a polar code and therefore is 
called the \emph{polar code construction}. 

The polar code construction is critical to obtain the best performance at finite 
blocklengths. As we mentioned, the polar code construction has an explicit 
definition in theory. It is challenging in practice because precise estimation 
of the bit-channels is intractable. Therefore, a wide range of approximate 
construction methods are proposed in \cite{ARI09tit, ARI08clett, TAL13, TRI12, 
MOR09clett, MOR09isit, ZHA14clett, LI13, WU14, KER14, BON12arxiv2, ZHA11}. 

An important characteristic of polar codes is their \emph{non-universality}. That is, 
different polar codes are generated depending on the specified value of 
signal-to-noise ratio (SNR), known as the \emph{design-SNR}. A change in operating 
SNR is possible in practice but a change in code according to SNR is not desirable. 
Therefore we wish to construct a polar code at one design-SNR and use it for a range 
of possible SNRs. As we see later, the choice of design-SNR is critical for the 
performance at all SNRs of interest. Unfortunately, there has been no study to 
identify the best design-SNR for any polar code construction.

A major issue with polar codes has been their inferior BER performance at finite 
blocklengths, compared to the state-of-the-art LDPC and Turbo codes of similar 
blocklength \cite{TAL11}. Several ideas have been proposed to overcome this 
issue. One may use a list decoder \cite{TAL11} to overcome this problem but this 
comes at the expense of an increase in decoding complexity.

In this paper, we aim at finding the best polar code construction algorithm among a 
range of available algorithms over a binary input additive white Gaussian noise 
(BI-AWGN) channel. We first propose a simple search algorithm to find the best 
design-SNR for each 
construction algorithm. Then we can find the best code construction algorithm among 
all. We find in our extensive simulations that all construction algorithms produce 
equally good polar codes when design-SNR is optimized.

The rest of the paper is organized as follows. In \cref{PC}, we present brief 
introduction to polar codes and outline our notation and channel model. In \cref{PC}, 
we describe the encoding, successive cancellation decoding and the construction of 
polar codes. In \cref{SEC3}, we review four main polar code construction algorithms 
which form the basis for several other variations. In \cref{UPC}, we discuss the 
non-universality of polar codes. Our discussion includes a proposal to fairly compare 
all construction algorithms. In \cref{pccalgs}, we give a detailed description along 
with pseudocode for efficient implementations of polar code constructions. We finally 
present our simulation results in \cref{SIMSEC} and conclude in \cref{CONCL}.

\section{Polar Codes}\label{PC}
Given any subset of indices $\mathcal{I}$ of elements of a vector 
$\b{x}$, we denote the corresponding sub-vector as $\b{x}_\mathcal{I}$. 
Similarly, when $\mathcal{I}$ denotes the indices of columns of a matrix 
$\b{A}$, the corresponding sub-matrix is denoted $\b{A}_{\mathcal{I}}$.

A \emph{polar code} may be specified completely by $(N,K,\mathcal{F})$, where $N$ is 
the length of a \emph{code word} in bits, $K$ is the number of information bits 
encoded per codeword, and $\mathcal{F}$ is a set of $N-K$ integer indices called 
\emph{frozen bit locations} from $\{0,1,\ldots,N-1\}$.

\noindent\emph{Encoding ---} For an $(N,K,\mathcal{F})$ polar code we describe below 
the encoding operation for a vector of information bits $\b{u}$ of length $K$. 
The \emph{rate} of the code is $R=K/N$. Let 
$n \triangleq \log_2(N)$ and $\b{F}^{\otimes n}= \b{F} \otimes \cdots \otimes \b{F}$ 
($n$ copies) be the $n$-fold Kronecker product of Arikan's standard 
polarizing kernel $\b{F} \triangleq \left[\begin{smallmatrix} 1&1\\ 0&1 
\end{smallmatrix}\right]$.

Then, a codeword is generated as
\begin{equation}
\b{x} = \b{G}\cdot \b{u} = \left(\b{F}^{\otimes n}\right)_{\mathcal{F}^c}\cdot 
\b{u},  \label{encodingeq}
\end{equation}
where $\mathcal{F}^c \triangleq  \{0,1,\ldots,N-1\}\backslash \mathcal{F} $
corresponds to the set of non-frozen bit indices and $\b{G}\triangleq 
\left(\b{F}^{\otimes n}\right)_{\mathcal{F}^c}$ is the \emph{generator matrix} of 
polar code.
A compact alternative form is
\begin{equation}
  \b{x} = \b{F}^{\otimes n} \b{d} \label{encodingeq_simple}
\end{equation}
where $\b{d} \in \{0,1\}^N$ is such that $\b{d}_\mathcal{F}
=0$ and $\b{d}_{\mathcal{F}^c} = \b{u}.$

\begin{figure}
\centering
\begin{tikzpicture}
\node at (0,0) {\includegraphics[width=2.8in,clip=true,trim=9mm 4mm 8mm 8mm] 
{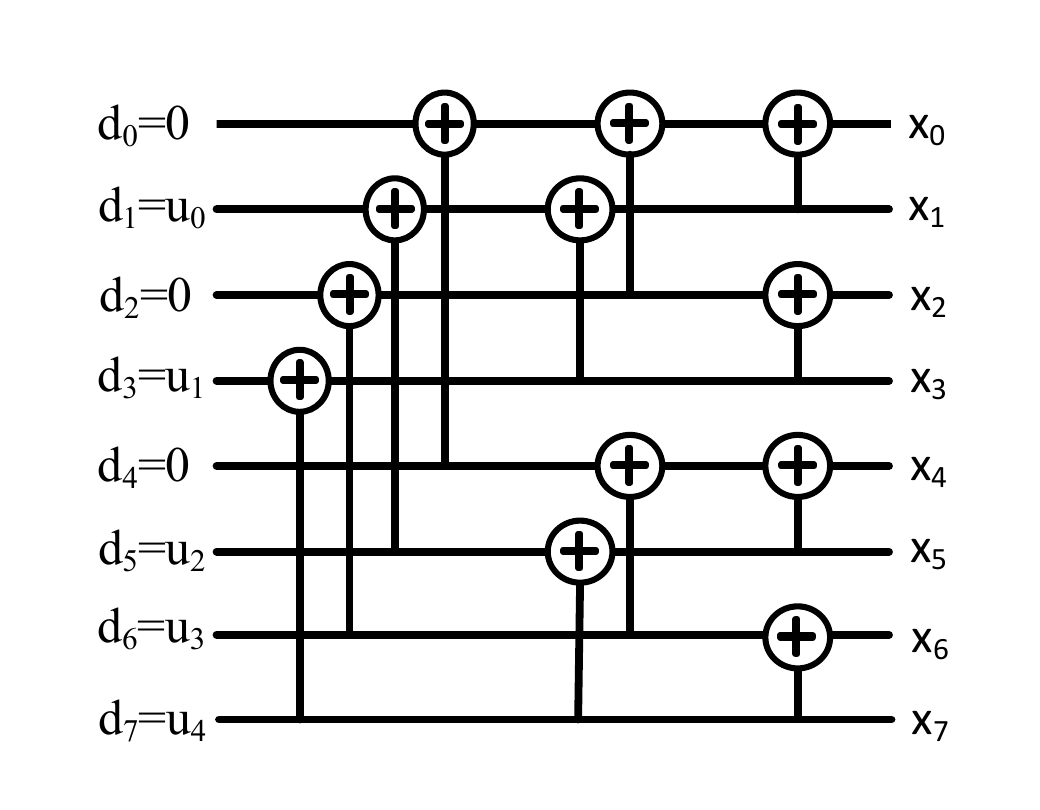}};
\node at (0,3.1) {$(N,K,\mathcal{F})=(8,5,\{0,2,4\})$};
\end{tikzpicture}
 \caption{\!Illustration of Arikan's $O(N\log_2N)$ complexity encoder implementation
of \eqref{encodingeq_simple} with $(N,K,\mathcal{F})=(8,5,\{0,2,4\})$\vspace{-1mm}}
\label{encoder_diag}
\end{figure}

Note that $\b{d}_\mathcal{F}$ is the set of \emph{frozen bits} as defined in 
Arikan's original formulation \cite{ARI09tit} and is taken here as zeros. Arikan 
also proposed an efficient implementation of complexity $\O{N\log N}$ of the 
encoding \eqref{encodingeq_simple} as shown in Fig.~\ref{encoder_diag}.

\noindent\emph{Modulation and Channel Model --- } Our channel is a BI-AWGN channel, 
with zero mean and variance $\tfrac{N_0}{2}$. Bits in $x$ are modulated as 
$\tilde{x}$ using
binary phase shift keying (BPSK). It maps $0\to -\sqrt{RE_b}$ and $1\to 
+\sqrt{RE_b}$, where $E_b$ denotes the energy spent per each information bit. We 
thus obtain the following channel
\begin{equation}
 y=\tilde{x}+n. \label{CHANNEL}
\end{equation}

Without loss of generality, we normalize the noise variace to be unity for the AWGN 
in all our future discussions.

\noindent\emph{Successive Cancellation Decoder (SCD) ---} The SCD algorithm
\cite{ARI09tit} essentially follows the same encoder diagram in 
Fig.\ref{encoder_diag} using decoding operations that resemble one iteration of 
the classic belief propagation algorithm. The likelihoods evolve in the reverse 
direction from right-to-left, using a pair of likelihood transformation equations, 
as illustrated with an example in \cite{ARI09tit}. Then the bit decisions are made 
at the left end of the circuit and broadcasted to the rest of the circuit. A 
complete pseudocode for implementing an SCD is available in \cite{HAR14isita}. The 
overall complexity is only $\O{N\log_2 N}$.

\noindent\emph{The Polar Code Construction ---}
The choice of the set $\mathcal{F}$ is a critical step in polar coding (i.e. the 
\emph{polar code construction}). This corresponds to the selection of 
best $K$ bit-channels among $N$, in terms of the bit error rate (BER) at a given 
value of $(RE_b/N_0)$ defined as the \emph{design-SNR}.

Since the exact BER of bit-channels is intractable, several approximations are 
used. This leads to many different constructions reviewed in \cref{SEC3}. The 
detailed algorithms are given later in \cref{pccalgs}.

\section{Overview of Current Literature on \\ Polar Code Constructions} 
\label{SEC3}

The earliest construction of polar codes is based on the evolution of simple bounds 
on the Bhattacharyya parameters of bit channels\cite{ARI08clett}. Though these bounds 
are proved in \cite{ARI09tit} only for BI-DMS channels, they may also be extended for 
infinite alphabet channels such as BI-AWGN. Although the bounds are loose for many of 
the bit-channels and more accurate methods are devised later, the codes designed 
using such bounds exhibit good performance. This construction enjoys the least 
$\O{N}$ complexity  among all (excluding the selection of $K$ best 
among $N$ metrics obtained). This includes $2N-1$ transformation operations 
corresponding to the 
transformation of upper bounds by polarizing kernel. This well-known method was also 
studied in a recent paper \cite{ZHA14clett}.

Another earlier construction is proposed in \cite{ARI09tit}, based on a Monte-Carlo 
simulation of the bit-channels. This can be applied to a wide range of channels 
including finite and infinite alphabet channels such as AWGN. However, the algorithm 
has the greatest complexity $\O{MN\log N}$ among all, where $M$ is the number of 
iterations of the Monte-Carlo simulation.

A more recent construction algorithm was proposed by Tal and Vardy \cite{TAL13} based 
an earlier proposal from Mori and Tanaka \cite{MOR09clett, MOR09isit}. At first it 
was proposed to find the bit channels by evaluating of their full finite alphabet 
distributions. The algorithm becomes intractable due to the explosion of the 
alphabet size to a power of $N$ by the end of $n$ channel transformations. This 
problem is specifically addressed in \cite{TAL13} by employing a novel low 
complexity close-to-optimal quantizer. In addition, they provide theoretical 
guarantees for the loss of performance due to the quantization. Note that, some 
channels are better estimated by simple bounds on Bhattacharyya parameters 
\cite{ARI08clett, ARI09tit}. Hence, in \cite{TAL13} the authors conclude a 
final algorithm by improving the BER estimates with the Bhattacharyya parameters 
whenever they are better. The final algorithm is considered by far the most accurate 
construction algorithm available with theoretical guarantees.

The algorithm is extendable to infinite output channels by using a quantization 
algorithm. In \cite{TAL13}, the authors propose a quantization algorithm for 
AWGN channels. When the channel output has $\mu$ symbols, the final complexity of the 
algorithm is $\O{N\cdot\mu^2\log \mu}$(excluding the selection of $K$ best 
among $N$ metrics obtained). This contains $2N-1$ number of $\O{\mu^2}$ 
complexity bit channel convolutions plus $2N-1$ number of $\O{\mu^2 \log \mu}$ 
complexity quantizer operations. The quantizer uses an intelligent data-structure 
that combines a heap and a list. The initialization of the algorithm involves the 
AWGN channel quantization to $\mu$ symbols, which takes an additional $\O{\mu}$ 
complexity. Overall, this algorithm has the second largest complexity, only next to 
the earlier Monte-Carlo based construction algorithm by Arikan.

For AWGN channels, the estimation of bit-channels based on Gaussian approximation is 
proposed in \cite{TRI12}. This enables to use the Gaussian distribution 
approximations on the intermediate likelihoods. This was found to well-approximate 
the actual bit-channels of polar codes \cite{LI13,WU14}. A similar algorithm after 
the original proposal in \cite{TRI12} was studied in \cite{LI13,WU14}.

The Gaussian approximation algorithm takes a complexity of $\O{N}$ function 
computations (excluding the selection of $K$ best 
among $N$ metrics obtained) similar to the 
Bhattacharyya bounds based algorithm, but involves relatively higher complexity 
function computations. Overall, this construction algorithm enjoys the second least 
complexity.

There also exist several other heuristic constructions, extended or inspired from the 
above constructions, e.g. \cite{KER14, BON12arxiv2, ZHA11}. However, 
these methods are less interesting due to poorer performance, higher complexity, 
and having no theoretical guarantees. Finally, constructions for different channels 
are also available in \cite{GOE13, BRA13, CHE13, HAS14}, for different input 
alphabets \cite{CAY12}, different kernels \cite{ZHA12wcsp, MIL12, SER14} and 
concatenated codes \cite{TRI11, MAH13isit}, which fall out of the scope of this 
paper.

\section{Non-Universality of Polar Codes}\label{UPC}
In coding theory, most of the codes are \emph{universal} in the sense that their 
definition is independent of the channel SNR, but polar codes are different. Arikan 
defines the set $\mathcal{F}$ of polar codes such that the block error rate (BLER) of 
polar codes is minimum under SCD. Since BLER is a function of SNR, it is not very 
surprising that the polar code changes with the given \emph{design-SNR}. Later in 
our simulations, we see that the change is very 
significant in terms of performance. There are a few recent attempts to design 
universal polar codes \cite{HAS14, ALS14itw, SAS14} but they come at a cost of much 
higher complexity at decoder and/or encoder. Further development of the theory of 
universal polar codes is required for their practical significance. In this paper, 
we restrict our attention towards Arikan's original polar codes only. We aim to 
design a polar code at a particular design-SNR and use it for a range of SNRs due to 
the following reasons.
\begin{enumerate}\itemsep=1.5mm
 \item In a number of experiments it is evident that the performance of polar codes 
constructed at one \emph{design-SNR} is good for a range of SNRs (see 
for e.g. \cite{TAL11}).
 \item The construction algorithms are not optimizing the performance 
exactly at the design-SNR.  That is, better performance at a given SNR may be 
obtained by constructing the code at a slightly different design-SNR (see 
\cref{SIMSEC}). This 
means even if we update the code dynamically with SNR, the performance may not be 
optimal.
\end{enumerate}

As a result, the problem reduces to finding at what design-SNR, we should design the 
polar code. Unfortunately, to the best of our knowledge, there is no such study of 
polar codes in this direction. Many research works often consider a heuristic choice 
of design-SNR. 

In \cref{SIMSEC} we see that the choice of design-SNR is indeed critical for its 
performance at all SNRs. Further, we find that such a performance depends on many 
parameters such as rate, blocklength and the algorithm used for the construction. 
This observation makes an exhaustive search for the design-SNR inevitable. The 
following simple search algorithm is proposed to find the best design-SNR.

\begin{enumerate}\itemsep=1.5mm
 \item Consider a set of SNRs $\{S_1,$ $S_2,$ $\ldots,$ $S_m\}$ that 
covers the range of SNRs of interest.
 \item Design $m$ polar codes at design-SNRs equal to $S_i$, $i=1,\ldots,m$ 
(i.e. find an $\mathcal{F}$ at each $S_i$), using any given construction 
method.
 \item Plot the performance curves BER vs. SNR or BLER vs. SNR of all above polar 
codes.
 \item Select the curve that best suits the needs of the target application and 
declare the corresponding SNR for its construction.
\end{enumerate}

We may use the search for different construction algorithms and make a fair 
comparison of performance among all. Extensive simulations of this comparison 
strategy are presented in \cref{SIMSEC}.

\section{The Polar Code Construction Algorithms}\label{pccalgs}
In this section we review all four important polar code construction methods
denoted \textbf{PCC-0}, \textbf{PCC-1}, \textbf{PCC-2}, \textbf{PCC-3},
and provide full pseudocode implementations. The corresponding bit-channel metrics 
generated from each algorithm are denoted $\b{z}^{(0)}$, $\b{z}^{(1)}$, 
$\b{z}^{(2)}$, $\b{z}^{(3)}$ respectively.

We should use \emph{logarithmic domain} (or \emph{log-domain} in short) calculations 
especially at high blocklengths such as $N\geq 256$ to avoid underflow. For 
simplicity we mention linear domain formulas only.

\subsection{{\normalfont\textbf{PCC-0}}: Arikan's Bhattacharyya bounds of bit 
channels} \label{PCC-0}
This earliest construction is from Arikan \cite{ARI08clett,ARI09tit}, using 
Bhattacharyya parameters. He proved in \cite[Appendix-D]{ARI09tit} that a pair of 
upper bounds on the Bhattacharyya parameters of bit-channels evolve as simply as 
$\{z,z\} \to \{2z-z^2, z^2\}$ at each polarizing transform $\b{F}$. Due to its 
simplicity, this construction has been widely used, and produced good polar codes.

There is an important modification to be used over \cite{ARI08clett}, which is due 
to the non-universality of polar codes as explained in \cite{LI13}. The original 
recursive algorithm requires an initial value and this was proposed by Arikan as 
$0.5$\cite{ARI08clett}, corresponding to the worst BER. This initial is actually 
the Bhattacharyya parameter of the underlying BI-AWGN channel, therefore it may be 
replaced with $\exp(-RE_b/N_0)$\cite{LI13}. Now, the original initial $0.5$ will be 
obtained at $\tfrac{RE_b}{N_0}=-1.5917$dB. The final algorithm is given 
below as Algorithm \textbf{PCC-0}.

A sample run of \textbf{PCC-0} reveals that some of the bounds quickly increase and 
may even increase beyond $0.5$ (but always $<1$). This suggests a very inaccurate 
channel estimation, which motivated several alternate constructions. However, the 
advantage of the alternate constructions was not characterized well in literature.

\IncMargin{1.8ex}
\RestyleAlgo{boxruled}
\SetAlCapSkip{0mm}
\SetAlgoSkip{}
\SetAlgoInsideSkip{}
\setlength{\interspacetitleboxruled}{2mm} %before & after the caption/title till 
% hlines.
\begin{algorithm}[!h] \footnotesize %\small
\linespread{1.3}\selectfont  %line spacing adjustment 
% tex.stackexchg.com/questions/138976
\DontPrintSemicolon
%%%%%%%%%%%%%%%%%%%%%%%%%%%%%%%%%%%%%%%
\SetAlgorithmName{Algorithm}{}{}
\SetAlgoRefName{PCC-0} %no number to be used to refer and label in caption.
%%%%%%%%%%%%%%%%%%%%%%%%%%%%%%%%%%%%%%%
\SetAlgoCaptionSeparator{}
\SetAlgoLined

\SetNlSty{}{}{:} %line numbering style: {font}{before-number}{after-number}
\SetNlSkip{3mm} %spacing from line number to text.

\SetKwComment{Comment}{\footnotesize$\vartriangleright\;$}{} %left marker, 
% right marker
\SetSideCommentRight %for inline comments, flushes to right
\SetKwInOut{Input}{INPUT}
\SetKwInOut{Output}{OUTPUT}

\Input{$N$, $K$, and design-SNR $EdB = (RE_b/N_0$ in dB)}
\Output{$\mathcal{F} \subset \{0,1,\ldots,N-1\}$ with $|\mathcal{F}|=N-K$}
\BlankLine
% $S=\tfrac{2K}{N}10^{EdB/10}$\;
$S=10^{EdB/10}$ and $n= \log_2N$\;
$\b{z}^{(0)} \in \mathbb{R}^N$, initialize $\b{z}^{(0)}[0]=\exp{(-S)}$ \;
\For(\Comment*[f]{\normalfont\footnotesize for each stage in 
\cref{encoder_diag}, right-to-left}){$j=1:n$}{
$u=2^{j}$\; %\Comment*[r]{\normalfont\footnotesize the sub-stage size}
     \For(\Comment*[f]{\normalfont\footnotesize For each connection}){$\ t = 
0:\tfrac{u}{2}-1~$
}{
$T=\b{z}^{(0)}[t]$\;
$\b{z}^{(0)}[t] = 2T - T^2$ \Comment*[r]{\normalfont\footnotesize Upper channel}
$\b{z}^{(0)}[u/2 + t] = T^2$ \Comment*[r]{\normalfont\footnotesize Lower channel}
}
}
$\mathcal{F} = \b{indices\_of\_greatest\_elements}\Big(\b{z}^{(0)},N-K\Big)$ 
\newline {\normalsize //} \normalfont\footnotesize Find indices of the greatest 
$N-K$ elements\; \vspace{2mm}
% $[\b{z}^{(0)}, idx] = \mathbf{Sort}(\b{z}^{(0)},\text{'descending'})$ 
% \Comment{\normalfont\footnotesize obtain in $idx$, the indices of $\b{z}^{(0)}$ 
% vector when sorted in descending order} \vspace{2mm}
% $\mathcal{F} = idx[0:N-K-1]$~~~~~~ \Comment{\normalfont\footnotesize Store 
% the first $N-K$ indices}
Return $\mathcal{F}$
\caption{: {The Bhattacharyya bounds}}
\end{algorithm}

\RestyleAlgo{boxruled}
\setlength{\interspacetitleboxruled}{1mm} %before & after the caption/title 
% till hlines.
\begin{algorithm}[!h] \footnotesize %\small
\linespread{1.1}\selectfont  %line spacing adjustment, Not found in manual
% from stackexchange page:  http://tex.stackexchange.com/questions/138976
% Alternatively, add \usepackage{setspace} and say \setstretch{1.35} in the 
% same place.

\DontPrintSemicolon
\SetAlgorithmName{Function}{}{}
\SetAlgoRefName{} %no number to be used to refer and label in caption.
\SetAlgoCaptionSeparator{~}
\SetAlgoLined

\SetNlSty{}{}{:} %line numbering style: {font}{before-number}{after-number}
\SetNlSkip{3mm} %spacing from line number to text.

\SetKwComment{Comment}{\footnotesize$\vartriangleright\;$}{} %left marker, 
	% (remember C-style /* ... */)
	% The right triangle marker \vartriangleright (as in algorithmic,
	% algorithmicx packages)requires "amssymb"
\SetSideCommentRight %for inline comments, flushes to right
\SetFillComment %for separate line comments, flushes to right

\SetKwInOut{Input}{INPUT}
\SetKwInOut{Output}{OUTPUT}
\BlankLine
\Input{~Vector $\b{v}$ of dimension $|\b{v}|\times 1$ and integer $l$ }
\Output{~$\mathcal{I}$,~ an $l\times 1$ integer vector containing $l$ indices in 
$\{0,1,\ldots,|\b{v}|-1\}$}
\BlankLine

$[\b{v}, idx] = \mathbf{Sort}(\b{v},\text{'descending'})$ \newline
//\normalfont\footnotesize obtain in $idx$, the $|\b{v}|$ indices of vector $\b{v}$ 
when sorted in descending order \vspace{1mm}\;
$\mathcal{I} = idx[0:l-1]$~~~~~~ \vspace{1mm}\Comment{\normalfont\footnotesize 
Store the first $l$ indices}
Return $\mathcal{I}$\vspace{2mm}\newline
// Note: This is a simple implementation which is okay for \newline
// $|\b{v}|$ up to a few thousands. For optimal performance, one should\newline
// use more advanced \emph{selection algorithms}\cite{FLO75,KIW05}.  \;

 \caption{\small$\b{indices\_of\_greatest\_elements(\b{v},~}l\b{)}$}
\label{FINDHIGHEST}
\end{algorithm}

\subsection{{\normalfont\textbf{PCC-1}}: Arikan's Monte-Carlo estimation of bit 
channels}\label{PCC-1}
A Monte-Carlo estimation of the bit-channel metrics is proposed in \cite{ARI09tit}. 
As a simulation based algorithm, it can be applied to a variety of channels.

We consider two specific improvements to the original proposal in \cite{ARI09tit} as 
follows.
\begin{enumerate}
 \item We simulate all-zero codeword transmission only, since the polar code is a 
linear code. Each iteration is now equivalent to an SCD with all bits treated 
as frozen.
 \item We calculate BER of bit-channels rather than their Bhattacharyya parameters 
\cite[Eq. (54) and (80)]{ARI09tit}.
\end{enumerate}

The first modification simplifies the algorithm by avoiding the encoding operation
for each simulated transmission, in addition to avoiding the step of updating the 
bit decisions in SCD. The second modification improves the precision of the 
estimate, 
simply because we use the exact BERs. This reflects in the overall BLER equation 
\cite[Eq. (3)]{WU14} which is relatively more accurate than \cite[Eq. (54) and 
(65)]{ARI09tit}. The complexity reduces to half due to these modifications. Also, as 
mentioned earlier the likelihood operations (step-\ref{LRTX1}, step-\ref{LRTX2} of 
$\b{UpdateL}$ and step-\ref{LRINIT},step-\ref{LRDECIDE} of \textbf{PCC-1}) are given 
in linear domain, but it is preferable to perform these operations in log-domain.

One particular disadvantage of this construction is the accuracy of the 
construction, 
which is limited severely by the Monte-Carlo iterations $M$. More specifically, the 
channels with BER less than or close to $1/M$ receive highly unreliable estimates. 
There exist many bit-channels that have extremely low  BER, due to the polarization 
effect. Most of such good channels will receive a zero estimate, which makes a 
comparison difficult.

When the rate is high enough, the construction works well, since the good channels 
are always chosen for the information transmission. This avoids the need of any 
comparison among the good channels. On the other hand, when the rate is very low, 
the choice will only be among channels that tend to be very good. In that case, any 
choice would result in approximately the same performance.

% \IncMargin{1.8ex} %Important for spacing between text and the box-lines all 4 
% sides around (visible or not). Especially **very important** for sufficient 
% spacing accommodation for line numbers.
% 	Its a one time setting for all subsequent algorithms, which adds the 
% specified amount to the default value. If used multiple times, then it gets added
% multiple times, moving margins further and further for each call (visible in 
% algorithms after each call).

\RestyleAlgo{boxruled}
\SetAlCapSkip{0mm}
\SetAlgoSkip{}
\SetAlgoInsideSkip{}
\setlength{\interspacetitleboxruled}{1mm} %before & after the caption/title till 
% hlines.
% \setlength{\interspacetitleboxruled}{1mm} %before & after the caption/title till 

\begin{algorithm}[!h] \footnotesize %\small
\linespread{1.2}\selectfont  %line spacing adjustment, Not found in manual
% from stackexchange page:  http://tex.stackexchange.com/questions/138976
% Alternatively, add \usepackage{setspace} and say \setstretch{1.35} in the same place.

\DontPrintSemicolon
\SetAlgorithmName{Algorithm}{}{}
\SetAlgoRefName{PCC-1} %no number to be used to refer and label in caption.
\SetAlgoCaptionSeparator{}
\SetAlgoLined

\SetNlSty{}{}{:} %line numbering style: {font}{before-number}{after-number}
\SetNlSkip{3mm} %spacing from line number to text.

\SetKwComment{Comment}{\footnotesize$\vartriangleright\;$}{} %left marker, right
	% (remember C-style /* ... */)
	% The right triangle marker \vartriangleright (as in algorithmic,
	% algorithmicx packages)requires "amssymb"
\SetSideCommentRight %for inline comments, flushes to right

\SetKwInOut{Input}{INPUT}
\SetKwInOut{Output}{OUTPUT}
\BlankLine

\Input{$N, K, $ the design-SNR $EdB = RE_b/N_0$ in dB,\\ $M=$ Monte-Carlo size, 
$randn()$ --- a standard \\ Gaussian pseudo random number generator}
\Output{$\mathcal{F} \subset \{0,1,\ldots,N-1\}$ with $|\mathcal{F}|=N-K$}
\BlankLine

Allocate and make visible to the other functions $N\times (n+1)$ matrices $\b{B}$, 
$\b{L}$ and set $\b{B}=0$\;
Bit-channel metrics $\b{c}\in\mathbb{R}^N$, initialize $\b{c}=0$\;
$n=\log_2N$ and $S=10^{EdB/10}$\; %$S=\tfrac{2K}{N}10^{EdB/10}$\;
\For{$t=1:M$}{
$\b{y}=-\sqrt{S}+randn(N,1)$ \Comment*[r]{\normalfont\footnotesize normalized 
channel, all-zero tx}
$\b{L}=\NaN$  \Comment*[r]{\normalfont\footnotesize Reset $\b{L}$}
Initialize the last column of $\b{L}$: \newline
$~~~\b{L}[j][n]=$Pr$(y_j|0)/$Pr$(y_j|1)=\exp{(-2y_j\sqrt{S})} ~~\forall j$ 
\label{LRINIT}\;
\For(\Comment*[f]{\normalfont The SCD with all frozen}) 
{$i=0,1,\ldots,N-1$}{
$l=\b{bitreversal}(i)$\;
$\b{UpdateL}(l,0)$ \Comment*[r]{\normalfont\footnotesize Update $\b{L}$, esp. 
$\b{L}[l][0]$}
%  $\b{B}[l][0]=0$ \Comment*[r]{\normalfont\footnotesize No error propagation}

 $\b{\hat{d}}[l]=\begin{cases}
                      0, \text{ if } \b{L}[l][0] \geq 1 \\
                      1, \text{ else }
                     \end{cases} $ \label{LRDECIDE}\;

%  $\b{UpdateB}(l,0)$  \Comment*[f]{\normalfont Broadcast bit-$l$ \&~update~$\b{B}$}
}
$\b{c}=\b{c}+\b{\hat{d}}$ \Comment*[r]{\normalfont\footnotesize Real addition}
}
$\b{z}^{(1)}=\b{c}/M$ \Comment*[r]{\normalfont\footnotesize Monte-Carlo averaging for 
BERs (not necessary)}
$\mathcal{F} = indices\_of\_highest\_elements\Big(\b{z}^{(1)},N-K\Big)$ 
\newline
{\normalsize //} \normalfont\footnotesize Find indices of the highest $N-K$ elements 
\; \vspace{2mm}
% $[\b{z}^{(1)}, idx] = \mathbf{Sort}(\b{z}^{(1)},\text{'descending'})$ 
% \Comment{\normalfont\small 
% obtain in $idx$, the indices of $\b{z}^{(1)}$ vector when sorted in descending 
% order}
% $\mathcal{F} = idx[0:N-K-1]$~~~~~~ \Comment{\normalfont\small Store the first $N-K$ 
% indices}
return $\mathcal{F}$
\caption{: The Monte-Carlo estimation}
\label{SCDalgm}
\end{algorithm}

\RestyleAlgo{boxruled}
\setlength{\interspacetitleboxruled}{1mm} %before & after the caption/title 
% till hlines.
\begin{algorithm}[!h] \footnotesize %\small
\linespread{1.1}\selectfont  %line spacing adjustment, Not found in manual
% from stackexchange page:  http://tex.stackexchange.com/questions/138976
% Alternatively, add \usepackage{setspace} and say \setstretch{1.35} in the 
% same place.

\DontPrintSemicolon
\SetAlgorithmName{Function}{}{}
\SetAlgoRefName{} %no number to be used to refer and label in caption.
\SetAlgoCaptionSeparator{~}
\SetAlgoLined

\SetNlSty{}{}{:} %line numbering style: {font}{before-number}{after-number}
\SetNlSkip{3mm} %spacing from line number to text.

\SetKwComment{Comment}{\footnotesize$\vartriangleright\;$}{} %left marker, 
	% (remember C-style /* ... */)
	% The right triangle marker \vartriangleright (as in algorithmic,
	% algorithmicx packages)requires "amssymb"
\SetSideCommentRight %for inline comments, flushes to right
\SetFillComment %for separate line comments, flushes to right

\SetKwInOut{Input}{INPUT}
\SetKwInOut{Output}{OUTPUT}
\BlankLine

\Input{Element indices $i,j$}
\Output{Recursively updated matrix $\b{L}$}
\BlankLine

$u=2^{n-j}$ and $l= (i$ mod $u)$\;

\uIf(\Comment*[f]{\normalfont Upper branch}){$l<u/2$}
{
    \SetAlgoVlined
    \If{$(\b{L}[i][j+1]=\NaN)$} % \textbf{then} \;
{$\b{UpdateL}(i,j+1)$; \textbf{end};}
    \If{$(\b{L}[i+u/2][j+1]=\NaN)$}
{$\b{UpdateL}(i+u/2,j+1)$; \textbf{end};}
$\b{L}[i][j] = 
\dfrac{\b{L}[i][j+1]\b{L}[i+u/2][j+1]+1}{\b{L}[i][j+1]+\b{
L}[i+u/2][j+1
] } $ \label{LRTX1}
% \vspace{2mm}
}
\Else(\Comment*[f]{\normalfont Lower branch}){
     \uIf{$\b{B}[i-u/2][j]=0$}
{ $\b{L}[i][j] = \b{L}[i][j+1]\b{L}[i-u/2][j+1]$ }
     \Else{ $\b{L}[i][j] = 
\dfrac{\b{L}[i][j+1]}{\b{L}[i-u/2][j+1]}$ \label{LRTX2}}
}

 \caption{\small$\b{UpdateL(i,j)}$ ~:~ Recursive LR comp. of SCD}
\label{SC_LR_rec}
\end{algorithm}

\subsection{{\normalfont\textbf{PCC-2}}: Tal \& Vardy's estimation of bit-channel 
TPMs}\label{PCC-2}
Tal \& Vardy's construction algorithm \cite{TAL13}, attempts to estimate the full 
transition probability matrices (TPMs) of bit-channels, instead of only estimating 
BERs. The desired BERs may be estimated from the TPMs. In spite of this effort, some 
estimates may be looser than the simple Bhattacharyya bounds calculated in 
\textbf{PCC-0}. This motivates to combine these two methods and come up with a 
hybrid 
algorithm that gives better estimates of the bit-channels compared to the individual 
algorithms. The final algorithm is denoted as Tal \& Vardy's construction for polar 
codes \cite{TAL13}. Like before, we compute BERs of bit-channels instead of 
Bhattacharyya parameters for improved accuracy.

When the channel is symmetric we can consider only the half of the channel. 
So when we say that the output alphabet is of size $\mu$ and its TPM is of 
dimension $2\times \mu$, we are actually referring to a symmetric channel of $2\mu$ 
output symbols and its TPM of dimension $2\times 2\mu$.

The output size of bit-channels grows rapidly, therefore a quantizer algorithm 
is proposed to control its size. It will then be used whenever the alphabet size 
exceeds a threshold $\mu$. The quantization is performed such that the channel's 
BER is preserved and the capacity of the quantized channel is maximized. In later 
part, we discuss the following components of the overall construction. 
\begin{enumerate}
 \item Quantization of the AWGN channel to initialize the code construction.
 \item The bit-channel convolutions corresponding to each use of Arikan's kernel 
$\b{F}$.
 \item The main quantizer algorithm for controlling the size of the alphabet after 
each bit-channel convolution.
 \item The main construction algorithm utilizing several component algorithms
discussed before.
\end{enumerate}

\noindent\emph{Quantization of the AWGN channel ---} Given an AWGN channel with 
BPSK modulation as defined in \eqref{CHANNEL} (with normalization), we may quantize 
its infinite alphabet to size $\mu$ before we use Tal \& Vardy's construction 
algorithm. Such a quantization algorithm is given in \cite{TAL13}, and reviewed 
below. 

The quantization involves first finding $\mu$ consecutive semi-open intervals of the 
positive real line $\mathbb{R}^+$. As explained below, the intervals are the 
pre-images of $\mu$ partitions of uniform length of $[0,1]$ (the \emph{range} of 
instantaneous capacity function).

% The intervals correspond to $\mu$ quantized 
% channel symbols and are such that at the boundaries, the instantaneous capacity 
% function defined below, takes values $\{0,1/\mu,2/\mu,\ldots,1\}$. Finally we obtain 
% a $2\times \mu$ dimensional TPM with entries found by integrating the channel 
% distributions over the respective intervals.

The instantaneous capacity function $C[\lambda(y)]$ is given by:
\begin{equation}
 C[\lambda(y)] = 1 - \log_2\left(1+\lambda\right) + 
\frac{\lambda}{\lambda+1}\log_2{\lambda}; \label{INSTCAPA1}
\end{equation}
where
\begin{equation}
\lambda(y)\triangleq \frac{f_Y{(y|0)}}{f_Y{(y|1)}}.
\end{equation}
In an AWGN channel according to \eqref{CHANNEL} under normalization, the 
conditional densities are: 
$f_Y(\cdot|0)\sim \mathcal{N}\left(-\sqrt{\tfrac{2RE_b}{N_0}},1\right)$ and 
$f_Y(\cdot|1)\sim \mathcal{N}\left(\sqrt{\tfrac{2RE_b}{N_0}},1\right)$.
Therefore, 
\begin{equation}
 \lambda(y) = \exp{\left(-2y\sqrt{\tfrac{2RE_b}{N_0}}\right)}\label{AWGNLR}
\end{equation}

We then obtain $\mu+1$ consecutive points $(a_0=0)< a_1< \cdots < 
a_{\mu-1} <(a_\mu=\infty)$ by solving the following equation. These points $\{a_i\}$
will partition the entire positive real line into $\mu$ intervals that correspond
to $\mu$ quantized symbols, such that
% 
% \begin{tikzpicture}
%  \draw[|->] (0,0) -- (4,0) -- (7,0);
%  \node at (0,-0.3) {$a_0=0$};
%  \node at (7,-0.3) {$a_\mu=\infty$};
%  \node at (2.25,-0.3) {$\cdots$};
%  \node at (4,-0.3) {$\cdots$};
%  \node at (1.5,0) {$)[$};
%  \node at (1.5,-0.35) {$a_1$};
%  \node at (3,0) {$)[$};
%  \node at (3,-0.35) {$a_i$};
%  \node at (5,0) {$)[$};
%  \node at (5,-0.35) {$a_{\mu-1}$};
% \end{tikzpicture}
% 
% The boundaries are obtained by solving the equation 
\begin{equation}
C[\lambda(a_i)]=\tfrac{i}{\mu},~\forall i=1,\ldots,\mu-1
\end{equation}
Finally the entries of the $2\times \mu$ dimensional TPM $\b{P}=[p_{ij}]$ of the 
quantized channel (with an input alphabet $\{0,1\}$ and an output alphabet 
$\{0,1,\ldots,\mu-1\}$) are obtained by integrating the corresponding channel 
distribution over the respective intervals as below.
\begin{align}
&p_{ij} \triangleq \Pr (j \text{ is received } ~|~i \text{ is 
transmitted})\nonumber \\[2mm]
&= \int_{a_{j}}^{a_{j+1}}f_Y(y|i)\;dy \nonumber \\[2mm]
&=\begin{cases}
Q\left(\sqrt{\tfrac{2RE_b}{N_0}}+a_j\right) - 
Q\left(\sqrt{\tfrac{2RE_b}{N_0}}+a_{j+1}\right), \text{if } i=0\\[2mm]
Q\left(-\sqrt{\tfrac{2RE_b}{N_0}}+a_{j}\right) - 
Q\left(-\sqrt{\tfrac{2RE_b}{N_0}}+a_{j+1}\right), \text{if } i=1
   \end{cases}\label{TPMDEF}
\end{align}
and 
\[
Q(x)\triangleq \tfrac{1}{\sqrt{2\pi}}\int_x^\infty\exp(-x^2/2)dx
\]
A full pseudocode implementation of the overall quantization of AWGN channel to 
$\mu$ symbols is given below.

\RestyleAlgo{boxruled}
\SetAlCapSkip{0mm}
\SetAlgoSkip{}
\SetAlgoInsideSkip{}
\setlength{\interspacetitleboxruled}{2mm} %before & after the caption/title till 
% hlines.
\begin{algorithm}[!h] \footnotesize %\small
\linespread{1.3}\selectfont  %line spacing adjustment 
% tex.stackexchg.com/questions/138976
\DontPrintSemicolon
%%%%%%%%%%%%%%%%%%%%%%%%%%%%%%%%%%%%%%%
\SetAlgorithmName{Function}{}{}
\SetAlgoRefName{discretizeAWGN\large()} %no number to be used to refer and 
% label in caption.
%%%%%%%%%%%%%%%%%%%%%%%%%%%%%%%%%%%%%%%
\SetAlgoCaptionSeparator{}
\SetAlgoLined
\SetNlSty{}{}{:} %line numbering style: {font}{before-number}{after-number}
\SetNlSkip{3mm} %spacing from line number to text.
\SetKwComment{Comment}{\footnotesize$\vartriangleright\;$}{} %left marker, 
% right marker
\SetSideCommentRight %for inline comments, flushes to right
\SetKwInOut{Input}{INPUT}
\SetKwInOut{Output}{OUTPUT}

\Input{Quantization size $\mu$, design-SNR $EdB=RE_b/N_0$ in dB}
\Output{TPM $\b{P}$ of size $2\times \mu$}
\BlankLine
% $S=\tfrac{2K}{N}10^{EdB/10}$\;
$S=10^{EdB/10}$\;
$\lambda(y)\triangleq \exp\left(-2y\sqrt{2S}\right)$ and 
\Comment*[r]{\normalfont\footnotesize Eq. \eqref{AWGNLR}} 
$C(x)\triangleq 1-\log_2(1+x) + x\log_2(x)/(1+x)$ 
\Comment*[r]{\normalfont\footnotesize Eq. \eqref{INSTCAPA1}}
$a \in \mathbb{R}^N$, initialize $a[0]=0$ and $a[\mu]= \infty$ \;
\For{$j=1:\mu-1$}{
$a[j]= \text{solve}\{C(\lambda(y)) = j/\mu \}$
}
\For(\Comment*[f]{\normalfont\footnotesize 
Eq. \eqref{TPMDEF}}){$j=0:\mu-1$}{
$\b{P}[0][j] = Q\left(\sqrt{2S}+a[j]\right) - 
Q\left(\sqrt{2S}+a[j+1]\right)$\;
$\b{P}[1][j] = Q\left(-\sqrt{2S}+a[j]\right) - Q\left(-\sqrt{2S}+a[j+1]\right)$
}
Return $\b{P}$
\caption{~}
\end{algorithm}

\noindent\emph{The bit-channel convolutions ---}
Each recursive use of the basic kernel $\b{F}=\left[\begin{smallmatrix} 1&1\\ 0&1 
\end{smallmatrix}\right]$ generates two polarized channels from a pair of identical 
channels $\mathcal{W}:\mathcal{X}\to\mathcal{Y}$. All the channels are 
represented by TPMs of two rows.

The pair of polarized channels are obtained by two self convolution operators 
$\boxplus$ and $\boxtimes$, referred to as upper-convolution and lower-convolution, 
respectively ($\oplus$ denotes usual binary EX-OR). The corresponding definitions 
are below\cite{ARI09tit,TAL13}.
\begin{multline}
 \mathcal{W}\boxplus\mathcal{W}(y_1,y_2~|i) ~= 
\frac{1}{2}\Big\{\mathcal{W}(y_1|0)\mathcal{W}(y_2|0\oplus i) ~+ \\
\mathcal{W}(y_1|1)\mathcal{W}(y_2|1\oplus i)\Big\}, \forall y_1,y_2\in 
\mathcal{Y}
\end{multline}
\begin{multline}
 \mathcal{W}\boxtimes\mathcal{W}(y_1,y_2,b~|i) = 
\frac{1}{2}\mathcal{W}(y_1|i)\mathcal{W}(y_2|b\oplus i) \\ \forall y_1,y_2\in 
\mathcal{Y} \text{ and } b\in\{0,1\}
\end{multline}

Proper care must be taken in applying these convolutions, to have all columns with 
likelihoods $>1$ (symmetric half of the channel). Whenever this gets violated, we 
simply swap the two probability values in the column, which brings its symmetric 
symbol in place. Another issue is whenever there are two or more columns with 
the same LR, we merge them to one by simply adding the columns.

Finally, we can see that the size of output alphabet of both the new channels is 
$\O{\mu^2}$. This increase in alphabet size is indeed high, which soon becomes 
intractable as we apply this transformation recursively. Hence the quantizer 
algorithms are essential.

\RestyleAlgo{boxruled}
\SetAlCapSkip{0mm}
\SetAlgoSkip{}
\SetAlgoInsideSkip{}
\setlength{\interspacetitleboxruled}{2mm} %before & after the caption/title till 
% hlines.
\begin{algorithm}[!h] \footnotesize %\small
\linespread{1.3}\selectfont  %line spacing adjustment 
% tex.stackexchg.com/questions/138976
\DontPrintSemicolon
%%%%%%%%%%%%%%%%%%%%%%%%%%%%%%%%%%%%%%%
\SetAlgorithmName{Function}{}{}
\SetAlgoRefName{upperConvolve\large()} %no number to be used to refer and 
% label in caption.
%%%%%%%%%%%%%%%%%%%%%%%%%%%%%%%%%%%%%%%
\SetAlgoCaptionSeparator{}
\SetAlgoLined
\SetNlSty{}{}{:} %line numbering style: {font}{before-number}{after-number}
\SetNlSkip{3mm} %spacing from line number to text.
\SetKwComment{Comment}{\footnotesize$\vartriangleright\;$}{} %left marker, 
% right marker
\SetSideCommentRight %for inline comments, flushes to right
\SetKwInOut{Input}{INPUT}
\SetKwInOut{Output}{OUTPUT}

\Input{A $2\times \mu$ TPM $\b{P}$ of channel to get convolved}
\Output{TPM $\b{Q}$ of size $2\times \mu', 
~\mu'\leq \mu(\mu+1)/2$}
\BlankLine
Allocate $\b{Q}\in \mathbb{R}^{2\times \mu(\mu+1)/2}$ and initialize $idx=-1$\;
\For{$i=0:\mu-1$}{
$idx = idx+1$\;
$\b{Q}[0][idx] = (\b{P}[0][i]^2 + \b{P}[1][i]^2)/2$\;
$\b{Q}[1][idx] = \b{P}[0][i]\b{P}[1][i] $\;
\For{$j=i+1:\mu-1$}{
$idx = idx+1$\;
$\b{Q}[0][idx] = \b{P}[0][i]\b{P}[0][j] + \b{P}[1][i]\b{P}[1][j]$\;
$\b{Q}[1][idx] = \b{P}[0][i]\b{P}[1][j] + \b{P}[1][i]\b{P}[0][j]$\;
\SetAlgoVlined
\If{$\b{Q}[0][idx]<\b{Q}[1][idx]$}{
swap($\b{Q}[0][idx]$, $\b{Q}[1][idx]$); \textbf{end};
}
}
}
Merge the columns of $\b{Q}$ with same LR (add columns) \;
Return $\b{Q}$
\caption{~}
\end{algorithm}

\RestyleAlgo{boxruled}
\SetAlCapSkip{0mm}
\SetAlgoSkip{}
\SetAlgoInsideSkip{}
\setlength{\interspacetitleboxruled}{2mm} %before & after the caption/title till 
% hlines.
\begin{algorithm}[!h] \footnotesize %\small
\linespread{1.3}\selectfont  %line spacing adjustment 
% tex.stackexchg.com/questions/138976
\DontPrintSemicolon
%%%%%%%%%%%%%%%%%%%%%%%%%%%%%%%
\SetAlgorithmName{Function}{}{}
\SetAlgoRefName{lowerConvolve\large()} %no number to be used to refer and
% label in caption.
%%%%%%%%%%%%%%%%%%%%%%%%%%%%%%%
\SetAlgoCaptionSeparator{}
\SetAlgoLined
\SetNlSty{}{}{:} %line numbering style: {font}{before-number}{after-number}
\SetNlSkip{3mm} %spacing from line number to text.
\SetKwComment{Comment}{\footnotesize$\vartriangleright\;$}{} %left marker, 
% right marker
\SetSideCommentRight %for inline comments, flushes to right
\SetKwInOut{Input}{INPUT}
\SetKwInOut{Output}{OUTPUT}

\Input{A $2\times \mu$ TPM $\b{P}$ of channel to get convolved}
\Output{TPM $\b{Q}$ of size $2\times \mu', ~\mu'\leq \mu(\mu+1)$}
\BlankLine
Allocate $\b{Q}\in \mathbb{R}^{2\times \mu(\mu+1)}$ and initialize $idx=-1$\;
\For{$i=0:\mu-1$}{
$idx = idx+1$\;
$\b{Q}[0][idx] = \b{P}[0][i]^2/2$ \;
$\b{Q}[1][idx] = \b{P}[1][i]^2/2$ \;
$idx = idx+1$ \;
$\b{Q}[0][idx] = \b{P}[0][i]\b{P}[1][i] $\;
$\b{Q}[1][idx] = \b{Q}[0][idx] $\;
\For{$j=i+1:\mu-1$}{
$idx = idx+1$\;
$\b{Q}[0][idx] = \b{P}[0][i]\b{P}[0][j]$\;
$\b{Q}[1][idx] = \b{P}[1][i]\b{P}[1][j]$\;
\SetAlgoVlined
\If{$\b{Q}[0][idx]<\b{Q}[1][idx]$}{swap($\b{Q}[0][idx]$, $\b{Q}[1][idx]$); 
\textbf{end};}
$idx = idx+1$\;
$\b{Q}[0][idx] = \b{P}[0][i]\b{P}[1][j]$\;
$\b{Q}[1][idx] = \b{P}[1][i]\b{P}[0][j]$\;
\If{$\b{Q}[0][idx]<\b{Q}[1][idx]$}{swap($\b{Q}[0][idx]$, $\b{Q}[1][idx]$); 
\textbf{end};}
}
}
Merge the columns of $\b{Q}$ with same LR (add columns)\;
Return $\b{Q}$
\caption{~}
\end{algorithm}

\noindent\emph{The quantizer algorithm ---}
A critical part of the construction algorithm is to find a good quantization 
algorithm, which was given in \cite{TAL13}, as discussed below. 
This quantization algorithm requires a special data-structure which is a combination 
of a \emph{heap} and a \emph{linked list}. We call this data-structure a 
\emph{heaplist}.

A \emph{heaplist} is a data-structure that holds $L$ symbol probabilities 
(columns) from the TPM as a list and $L-1$ loss-of-capacity values as a heap. These 
values in heap result from several symbol-merger operations, defined below.

One main objective of the data-structure is to enable a low complexity $\O{\log L}$ 
operation for extracting the minimum of the values in heap, while simultaneously 
maintaining the \emph{list} of $L$ symbols in a sorted order of likelihoods. 

A symbol-merger operation is the critical component of this quantization algorithm, 
and is a part of the heaplist. This operation is allowed only on two consecutive 
symbols in the list. A \emph{merger} of two columns in the list replaces the two 
columns by a single column equal to their sum corresponding to one new symbol, 
simultaneously adjusting the heap accordingly. Each merger reduces the size of the 
heaplist by one element and changes the heaplist.

Overall, a heaplist is simply a data-structure providing three main 
operations described as follows.
\begin{itemize}
 \item $initialize\_heaplist()$: Given a $2\times L$ TPM, initialize the list part 
of the data-structure with its columns in the increasing order of likelihoods, 
all $\geq 1$. Involves a sorting operation with $\O{L\log L}$ complexity. The heap 
part of the data-structure is stored with the $L-1$ values of loss in capacity when 
two consecutive symbols in the list are merged. These $L-1$ values in heap are 
tightly attached to the first $L-1$ values in the list.
 \item $minloss\_index()$: Find the value of minimum loss-of-capacity, and return the
index of the column in the list attached to the minimum loss. The index indicates 
the optimal symbol-merger, which merges the column and its next.
 \item $merge\_at\_index()$: Perform the merger operation of the two elements in 
list, present at the given index and its next. This will result in the reduction of 
size of the heap and the list, by one element. One entry in the list gets updated 
and one entry gets removed. Accordingly, a few entries in the heap will be updated.
 \item Miscellaneous:\\
 $size()$ -- to know the number of columns of the TPM currently maintained within 
the 
heaplist.\\
 $TPM()$ -- to extract the TPM from heap, in a two-row matrix format.\\
 $has\_duplicates()$ -- true or false based on whether there are columns in list 
with 
same likelihood ratio.
\end{itemize}

\RestyleAlgo{boxruled}
\SetAlCapSkip{0mm}
\SetAlgoSkip{}
\SetAlgoInsideSkip{}
\setlength{\interspacetitleboxruled}{2mm} %before & after the caption/title till 
% hlines.
\begin{algorithm}[!h] \footnotesize %\small
\linespread{1.3}\selectfont  %line spacing adjustment 
% tex.stackexchg.com/questions/138976
\DontPrintSemicolon
%%%%%%%%%%%%%%%%%%%%%%%%%%%%%%%
\SetAlgorithmName{Function}{}{}
\SetAlgoRefName{Quantize\_to\_size\large()} %no number to be used to refer and label 
% in caption.
%%%%%%%%%%%%%%%%%%%%%%%%%%%%%%%

\SetAlgoCaptionSeparator{}
\SetAlgoLined

\SetNlSty{}{}{:} %line numbering style: {font}{before-number}{after-number}
\SetNlSkip{3mm} %spacing from line number to text.

\SetKwComment{Comment}{\footnotesize$\vartriangleright\;$}{} %left marker, 
% right marker
\SetSideCommentRight %for inline comments, flushes to right
\SetKwInOut{Input}{INPUT}
\SetKwInOut{Output}{OUTPUT}

\Input{TPM $\b{P}$ of size $2\times L$, quantization size $\mu<L$}
\Output{Quantized TPM of size $2\times \mu',\mu'\leq\mu$}

\BlankLine
Instantiate a heaplist object $H$\;
$H.initialize\_heaplist(\b{P})$\;
\While{$H.size()>\mu \text{ \normalfont or } H.has\_duplicates()$}{
$idx = H.minloss\_index()$\;
$H.merge\_at\_index(idx)$\;
}
Return $H.TPM()$
\caption{~}
\end{algorithm}

\noindent\emph{The Overall Implementation and Pseudocode ---}
The following few steps are involved in the main algorithm to generate $N$ 
symmetric bit-channel TPMs.
\begin{enumerate}\itemsep=2mm
 \item Initialize the list with one finite alphabet channel equal to the quantized 
version of the AWGN channel having alphabet size $\mu$.
 \item For each finite alphabet channel in the list, repeat the following:
 \begin{enumerate}\itemsep=1mm
 \item Perform upper and lower convolutions of the channel with itself to 
generate two new channels of higher size.
 \item Quantize the two resultant channels to the size $\mu$ and replace the one 
original channel.
 \end{enumerate}
  \item If the number of channels is less than $N$, then repeat step 2, otherwise go 
to step 4.
  \item Calculate the BER of each channel and compare with its Bhattacharyya 
parameter calculated using \textbf{PCC-0}. Declare the least value as the 
bit-channel metric.
\end{enumerate}
The pseudocode of the same is provided as algorithm \textbf{PCC-2}. This represents 
the full construction algorithm of Tal \& Vardy\cite{TAL13}, except that we compute 
exact-BER of the bit-channels instead of their Bhattacharyya bounds for the 
comparison. This improves the precision of bit-channel metrics before they are 
compared and the greatest $N-K$ values are selected.

\RestyleAlgo{boxruled}
\SetAlCapSkip{0mm}
\SetAlgoSkip{}
\SetAlgoInsideSkip{}
\setlength{\interspacetitleboxruled}{2mm} %before & after the caption/title till 
% hlines.
\begin{algorithm}[!h] \footnotesize %\small
\linespread{1.3}\selectfont  %line spacing adjustment 
% tex.stackexchg.com/questions/138976
\DontPrintSemicolon
%%%%%%%%%%%%%%%%%%%%%%%%%%%%%%%
\SetAlgorithmName{Algorithm}{}{}
\SetAlgoRefName{PCC-2} %no number to be used to refer and label in caption.
%%%%%%%%%%%%%%%%%%%%%%%%%%%%%%%
\SetAlgoCaptionSeparator{}
\SetAlgoLined

\SetNlSty{}{}{:} %line numbering style: {font}{before-number}{after-number}
\SetNlSkip{3mm} %spacing from line number to text.

\SetKwComment{Comment}{\footnotesize$\vartriangleright\;$}{} %left marker, 
% right marker
\SetSideCommentRight %for inline comments, flushes to right
\SetKwInOut{Input}{INPUT}
\SetKwInOut{Output}{OUTPUT}

\Input{$N$, $K$ and design-SNR $EdB = (RE_b/N_0$ in dB); ~~~~$\b{z}^{(0)}$ is 
vector of bit-channel metrics from \textbf{PCC-0}}
\Output{$\mathcal{F} \subset \{0,1,\ldots,N-1\}$ with $|\mathcal{F}|=N-K$}
\BlankLine
% $S=\tfrac{2K}{N}10^{EdB/10}$ \;
Allocate $channels$, an array of $N$ TPMs, each of dimension $2\times \mu$. 
~~$ch1,ch2$ are to hold intermediate TPMs. \;
$n=\log_2 N$\;
$channels[0]=discretizeAWGN(\mu,EdB)$ ~~~//Initialize\;
\For(\Comment*[f]{\normalfont\small for each stage in 
\cref{encoder_diag}, right-to-left}){$j=1:n$}{
$u=2^{j}$\; %\Comment*[r]{\normalfont\footnotesize the sub-stage size}
     \For{$\ t = 0:\tfrac{u}{2}-1~$
}{
$ch1 = \b{upperConvolve}(channels[t])$\;
$ch2 = \b{lowerConvolve}(channels[t])$\;
$channels[t] = \b{Quantize\_to\_size}(ch1,\mu)$ 
$channels[u/2 + t] = \b{Quantize\_to\_size}(ch2,\mu)$
}
}
Allocate $\b{z}^{(2)}$, a real vector dimension $N\times 1$\;
\For{$i=0:N-1$}{
$\b{z}^{(2)}[i]=0$\;
$\mu'=$ no. of columns in $channels[i]$\;
\For{$j=0:\mu'-1$}{
$\b{z}^{(2)}[i] = \b{z}^{(2)}[i]+(channels[i])[1][j]$
}
$\b{z}^{(2)}[i] = \min\left\{\b{z}^{(0)}[i],\b{z}^{(2)}[i]\right\}$\;
}
$\mathcal{F} = \b{indices\_of\_greatest\_elements}\Big(\b{z}^{(2)},N-K\Big)$ 
\newline
{\normalsize //} \normalfont\footnotesize Find indices of the greatest $N-K$ 
elements\; \vspace{2mm}
% $[\b{z}^{(2)}, idx] = \mathbf{Sort}(\b{z}^{(2)},\text{'descending'})$ 
% \Comment{\normalfont\small obtain in $idx$, the indices of $z$ vector when sorted 
% in descending order} \vspace{1mm}
% $\mathcal{F} = idx[0:N-K-1]$~~~~~~ \Comment{\normalfont\small Store the first 
% $N-K$ indices}
Return $\mathcal{F}$
\caption{: {The full TPM estimation}}
\end{algorithm}

\subsection{{\normalfont\textbf{PCC-3}}: Trifonov's Gaussian approximation of bit 
channels}\label{PCC-3}
Gaussian approximation for polar code construction was first proposed by Trifonov 
\cite{TRI12} and rediscovered in \cite{WU14,LI13}.

The basic idea is to estimate all the log-likelihood ratios (LLR's) at intermediate 
stages as Gaussian variables. This simplifies the analysis by requiring only to 
calculate the mean and variance of the LLRs at each stage of decoding. The bit 
channel metrics in this case are different. We estimate a bit-channel metric 
proportional to the argument of a Q-function, which represents its BER under 
Gaussian 
approximation\cite{WU14}. So while being sufficient for comparison, the metrics in 
$\b{z}^{(3)}$ are not BERs but are inversely proportional to the BERs of 
bit-channels unlike the metrics from other constructions. 

The following interpolation function and its inverse are essentially used to perform 
this construction (as in \cite{CHU01}). 
\begin{equation}
 \phi(x)\triangleq \begin{cases}
                    \exp(-0.4527x^{0.86}+0.0218) \text{ if }0<x\leq10\\[2mm]
                 \sqrt{\tfrac{\pi}{2}}(1-\tfrac{10}{7x})\exp(-x/4), \text{ if } x>10
                   \end{cases}
\end{equation}

Full pseudocode to implement this algorithm is given below. We may employ bisection 
method to find the inverse of the function $\phi(x)$ (we observed that for this 
special function, bisection method works faster than Newton-Raphson method).

% \IncMargin{1.8ex} %to be used only once for the whole tex file
\RestyleAlgo{boxruled}
\SetAlCapSkip{0mm}
\SetAlgoSkip{}
\SetAlgoInsideSkip{}
\setlength{\interspacetitleboxruled}{2mm} %before & after the caption/title till 
% hlines.
\begin{algorithm}[!h] \footnotesize %\small
\linespread{1.3}\selectfont  %line spacing adjustment 
% tex.stackexchg.com/questions/138976
\DontPrintSemicolon
%%%%%%%%%%%%%%%%%%%%%%%%%%%%%%%
\SetAlgorithmName{Algorithm}{}{}
\SetAlgoRefName{PCC-3} %no number to be used to refer and label in caption.
%%%%%%%%%%%%%%%%%%%%%%%%%%%%%%%
\SetAlgoCaptionSeparator{}
\SetAlgoLined

\SetNlSty{}{}{:} %line numbering style: {font}{before-number}{after-number}
\SetNlSkip{3mm} %spacing from line number to text.

\SetKwComment{Comment}{\footnotesize$\vartriangleright\;$}{} %left marker, 
% right marker
\SetSideCommentRight %for inline comments, flushes to right
\SetKwInOut{Input}{INPUT}
\SetKwInOut{Output}{OUTPUT}

\Input{$N$, $K$ and design-SNR $EdB = (RE_b/N_0$ in dB)}
\Output{$\mathcal{F} \subset \{0,1,\ldots,N-1\}$ with $|\mathcal{F}|=N-K$}
\BlankLine
% $S=\tfrac{2K}{N}10^{EdB/10}$ \;
$S=10^{EdB/10}$ and $n=\log_2N$\;
$\b{z}^{(3)} \in \mathbb{R}^N$, initialize $\b{z}^{(3)}[0]=4S$ \;
\For(\Comment*[f]{\normalfont\small for each stage in 
\cref{encoder_diag}, right-to-left}){$j=1:n$}{
$u=2^{j}$\; %\Comment*[r]{\normalfont\footnotesize the sub-stage size}
     \For{$\ t = 0:\tfrac{u}{2}-1~$}{
$T=\b{z}^{(3)}[t]$\;
$\b{z}^{(3)}[t] = \phi^{-1}\left(1-(1-\phi(T))^2\right)$ 
\Comment*[r]{\normalfont\footnotesize Upper channel}
$\b{z}^{(3)}[u/2 + t] = 2T$ \Comment*[r]{\normalfont\footnotesize Lower channel}
}
}
$\mathcal{F} = \b{indices\_of\_least\_elements}\Big(\b{z}^{(3)},N-K\Big)$ 
\newline
{\normalsize //} \normalfont\footnotesize Find indices of the least $N-K$ 
elements \; \vspace{2mm}
% $[\b{z}^{(3)}, idx] = \mathbf{Sort}(\b{z}^{(3)},\text{'ascending'})$ 
% \Comment{\normalfont\small obtain in $idx$, the indices of $\b{z}^{(3)}$ vector 
% when sorted in ascending order}
% $\mathcal{F} = idx[0:N-K-1]$~~~~~~ \Comment{\normalfont\small Store the first 
% $N-K$ indices}
Return $\mathcal{F}$
\caption{: {The Gaussian approximation}}
\end{algorithm}

\RestyleAlgo{boxruled}
\setlength{\interspacetitleboxruled}{1mm} %before & after the caption/title 
% till hlines.
\begin{algorithm}[!h] \footnotesize %\small
\linespread{1.1}\selectfont  %line spacing adjustment

\DontPrintSemicolon
\SetAlgorithmName{Function}{}{}
\SetAlgoRefName{} %no number to be used to refer and label in caption.
\SetAlgoCaptionSeparator{~}
\SetAlgoLined

\SetNlSty{}{}{:} %line numbering style: {font}{before-number}{after-number}
\SetNlSkip{3mm} %spacing from line number to text.

\SetKwComment{Comment}{\footnotesize$\vartriangleright\;$}{} %left marker, 
	% (remember C-style /* ... */)
	% The right triangle marker \vartriangleright (as in algorithmic,
	% algorithmicx packages)requires "amssymb"
\SetSideCommentRight %for inline comments, flushes to right
\SetFillComment %for separate line comments, flushes to right

\SetKwInOut{Input}{INPUT}
\SetKwInOut{Output}{OUTPUT}
\BlankLine
\Input{~Vector $\b{v}$ of dimension $|\b{v}|\times 1$ and integer $l$ }
\Output{An $l\times 1$ integer vector containing $l$ indices in 
$\{0,1,\ldots,|\b{v}|-1\}$}
\BlankLine

Return $\Big(\b{indices\_of\_greatest\_elements(-\b{v},~}l\b{)}\Big)$

 \caption{\small$\b{indices\_of\_least\_elements(\b{v},~}l\b{)}$}
\label{FINDLOWEST}
\end{algorithm}

\section{Simulations and Discussion}\label{SIMSEC}
\begin{figure}
\centering
 \includegraphics[width=3.45in,clip=true,trim=18mm 7mm 21mm 8mm]{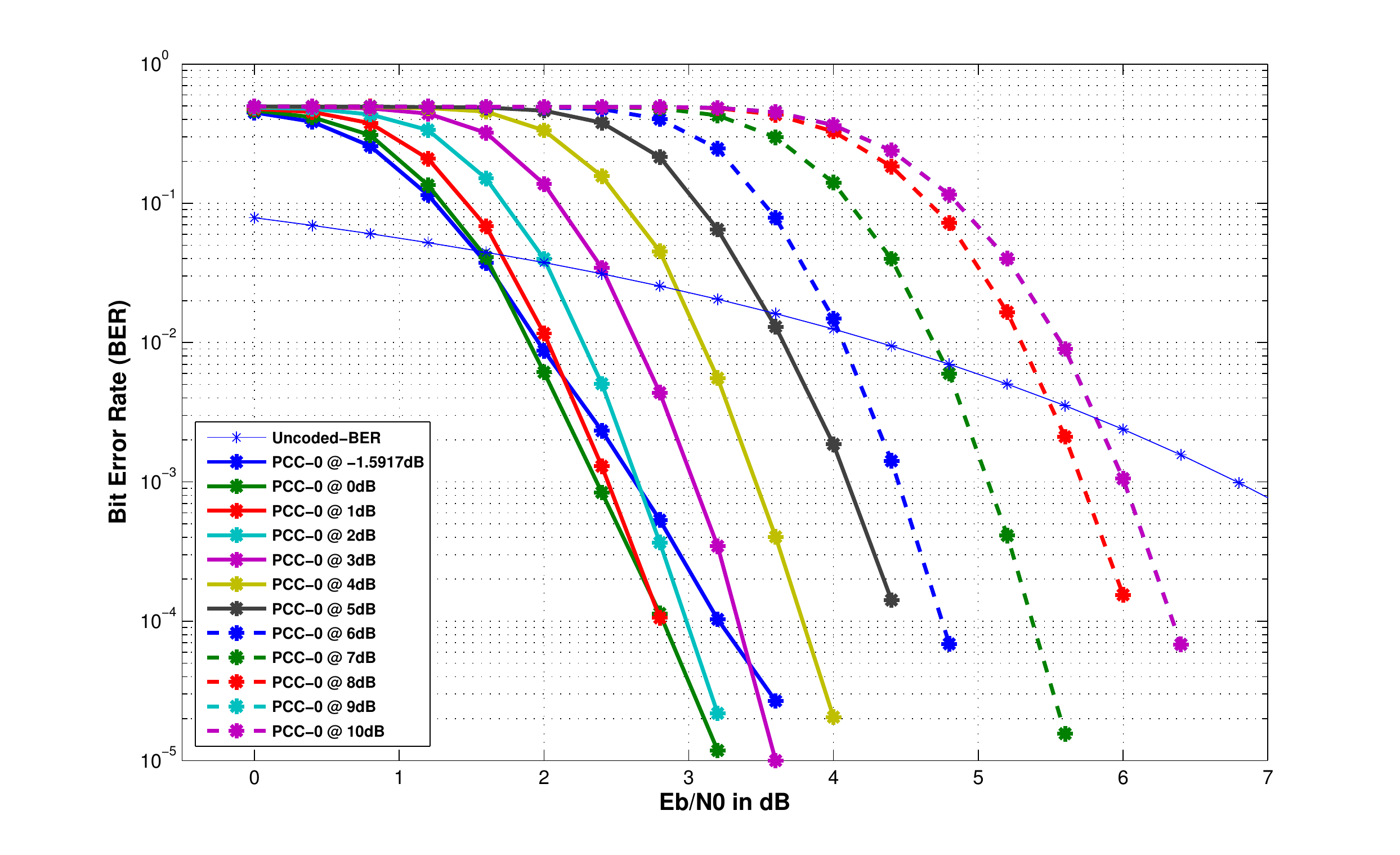}
 \caption{\textbf{PCC-0}: The effect of design-SNR at $N=2048$ and $R=0.5$ 
\vspace{-4mm}}
 \label{PCC0}
\end{figure}

\begin{figure}
\centering
 \includegraphics[width=3.45in,clip=true,trim=18mm 7mm 21mm 8mm]{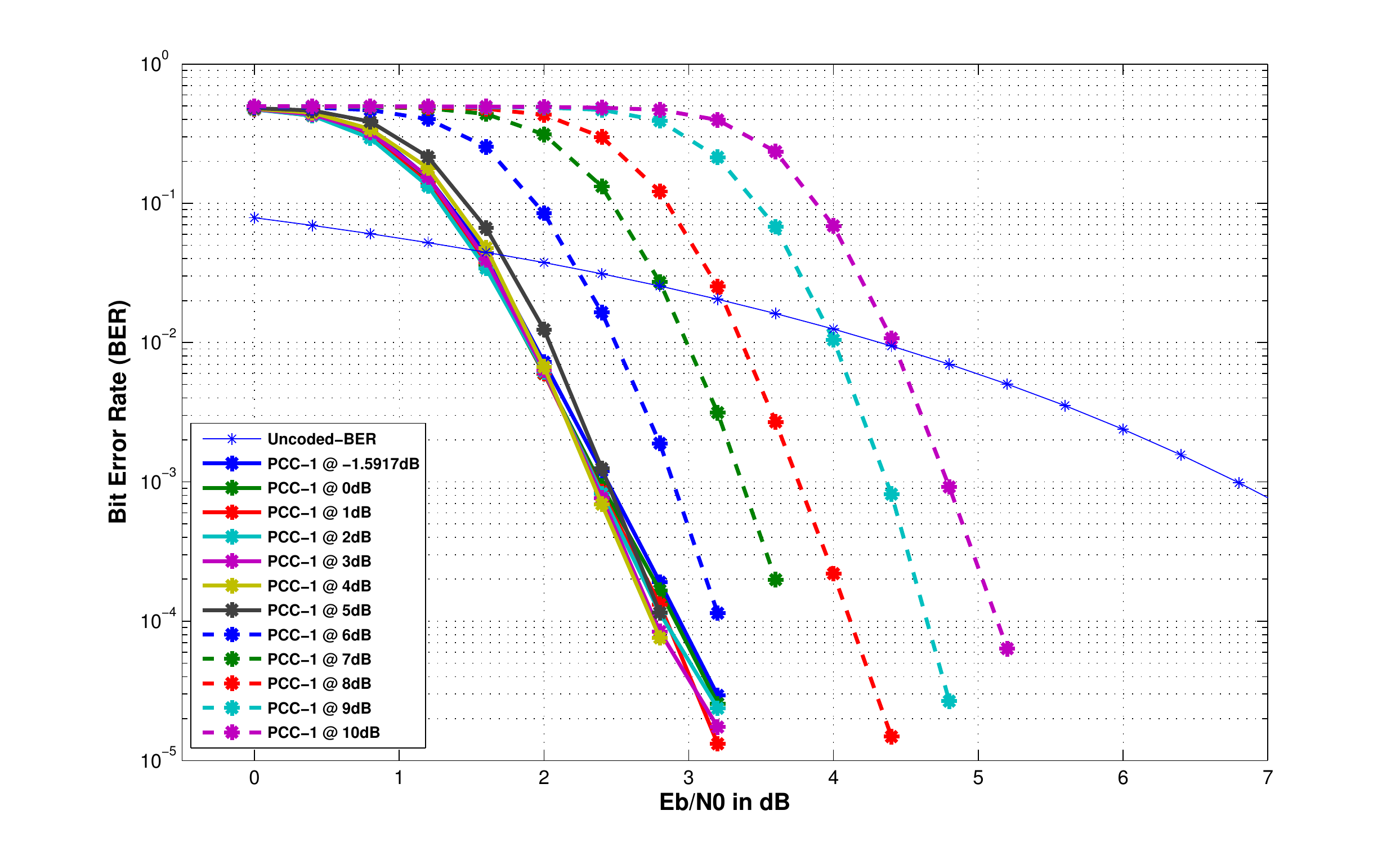}
 \caption{\textbf{PCC-1}: The effect of design-SNR at $N=2048$ and $R=0.5$ 
\vspace{-1mm}}
 \label{PCC1}
\end{figure}

\begin{figure}
\centering
 \includegraphics[width=3.45in,clip=true,trim=18mm 7mm 21mm 8mm]{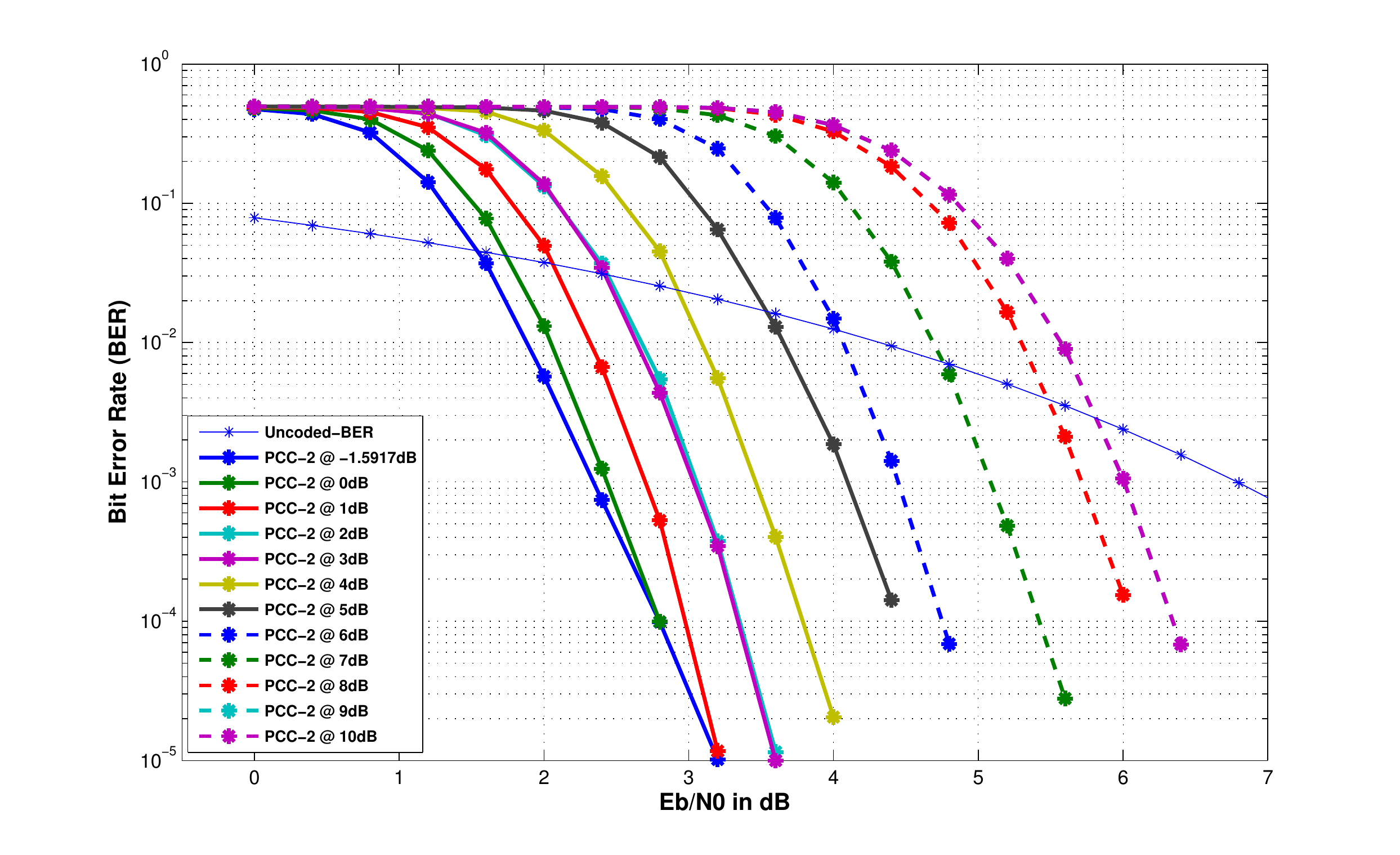}
 \caption{\textbf{PCC-2}: The effect of design-SNR at $N=2048$ and $R=0.5$ 
\vspace{-2mm}}
 \label{PCC2}
\end{figure}

\begin{figure}
\centering
 \includegraphics[width=3.45in,clip=true,trim=18mm 7mm 21mm 8mm]{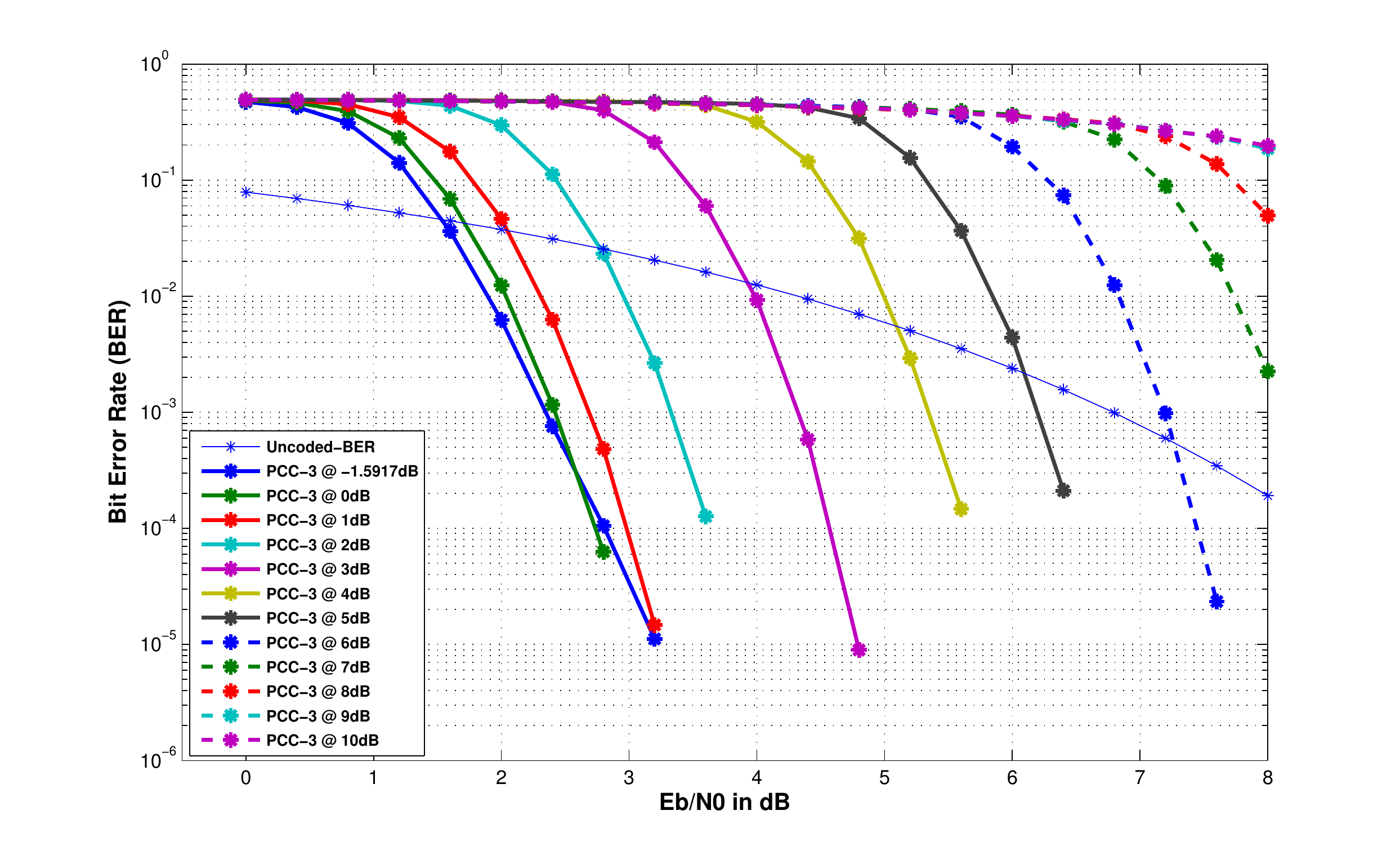}
 \caption{\textbf{PCC-3}: The effect of design-SNR at $N=2048$ and $R=0.5$ 
\vspace{-4mm}}
 \label{PCC3}
\end{figure}

In this section we consider comparing the performance of polar codes from various 
polar code constructions. \cref{PCC0,PCC1,PCC2,PCC3} present the performance 
of polar codes produced by each of algorithms \textbf{PCC-0} to \textbf{PCC-3} 
at $N=2048$ and $R=0.5$. Different curves in each figure represent the polar codes 
constructed at different design-SNRs. Clearly, design-SNR is critical for all 
construction algorithms to generate polars code with a good performance. 
Interestingly, \textbf{PCC-3} shows high performance variations with design-SNR.

We may observe that at high design-SNRs, the performance degrades with an 
increase in design-SNR for all the construction algorithms. In fact in the current 
example, \textbf{PCC-2} and \textbf{PCC-3} followed this more precisely and 
performed the best at the least design-SNRs. The least design-SNR we considered is 
$-1.5917$dB, at which \textbf{PCC-0} takes the worst-case initial $0.5$.

\begin{figure}
\centering
 \includegraphics[width=3.45in,clip=true,trim=18mm 7mm 21mm 8mm]{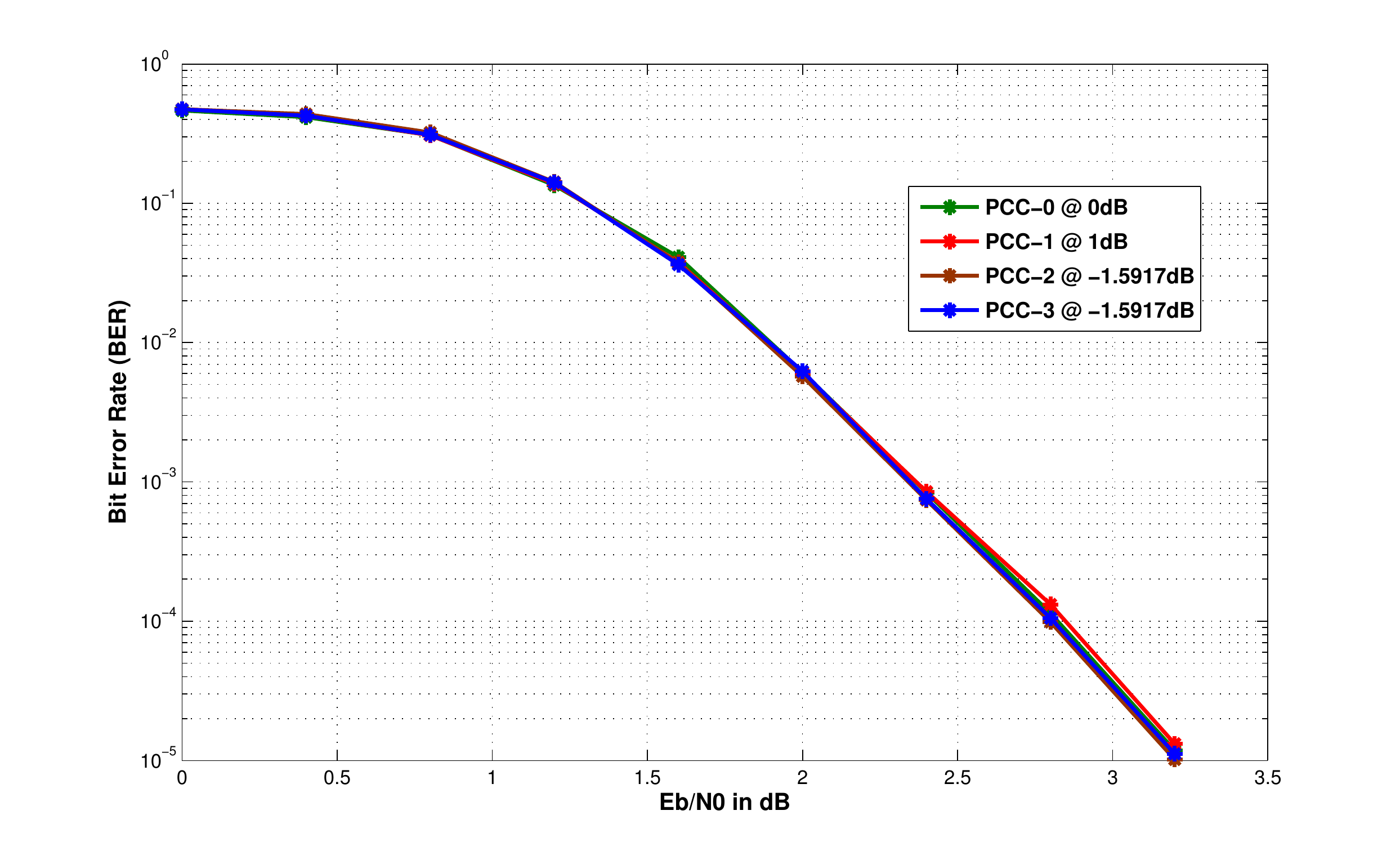}
 \caption{Comparison of all the constructions at their best design-SNRs and 
$N=2048$ and $R=0.5$}
 \label{PCC0123}
\end{figure}

As discussed in \cref{UPC}, the optimal design-SNR is a function of all possible 
parameters such as rate, blocklength and construction algorithm. We therefore select 
the best design-SNR for each construction algorithm from the above graphs and then 
make a fair comparison. This comparison is shown in \cref{PCC0123}. We see 
that if we can find the optimal design-SNR, any polar code construction algorithm 
produces polar codes of equally good performance. 


\section{Conclusions}\label{CONCL}
We have presented a comprehensive survey and full pseudocode implementations of all 
the well-known construction algorithms. We then proposed a simple discrete search 
to find the best design-SNR for any given polar code construction algorithm. We 
then compared various polar code constructions and concluded that all are equally 
good in AWGN if the design-SNR is optimized for the best performance. Thus in 
future, we may use simple algorithms only.

% \bibliographystyle{IEEEtran}
% \bibliography{allbibs_JCN.bib}

\end{document}